\newcommand{\be}{\begin{equation}}
\newcommand{\ee}{\end{equation}}
\def\bea{\begin{eqnarray}}
\def\eea{\end{eqnarray}}

\documentstyle[twoside,fleqn,espcrc2]{article}
\title{Neutrinos: Heralds of New Physics}

\author{P. Ramond\address{Institute for Fundamental Theory\\
Department of Physics\\ 
University of Florida\\ 
Gainesville, Fl 32611}}

\begin{document}

\begin{abstract}
\noindent The central role of neutrinos in the determination of fundamental interactions is reviewed. The recent SuperKamiokande discovery of neutrino mass gives an {\it aper\c cu} of physics at short distances, and tests theories of flavor.   
Quark-lepton symmetries, derived from  grand unification and/or string
theories, can help determine the  standard model parameters in the neutrino sector.

\end{abstract}
\maketitle
\section{A Neutrino Story}
Over the last seventy years, neutrinos have proved to be of central importance for  our understanding of fundamental interactions. From their existence which led to Fermi's theory of $\beta$ decay, to their chiral nature which held the key to parity viola
tion, and again today it is no surprise that  it is their masses 
that give the first indication of new interactions beyond the standard model.

Once it became apparent that the spectrum of $\beta$ electrons was 
continuous~\cite{CHAD,ELWO}, something drastic had to be done! 
 In December 1930,  in a  letter that starts with 
typical panache, $``${\it  Dear Radioactive Ladies and Gentlemen...}", 
W. Pauli puts forward a ``{\it desperate}" way out: there is a
companion neutral particle to the $\beta$ electron. Thus earthlings became aware of
the {\it neutrino}, so named in 
1933 by Fermi (Pauli's original name, {\it neutron},
superseded by Chadwick's discovery of a heavy neutral particle), 
implying that there is something small about it, specifically 
its  mass, although nobody at that time thought it was {\it that} small. 

Fifteen years later, B. Pontecorvo~\cite{PONTA}
proposes the unthinkable, that neutrinos can be detected:  an electron 
neutrino that hits a ${^{37}Cl}$ atom will transform it into the inert 
radioactive gas ${^{37}Ar}$, which  can be stored and then
detected through  radioactive decay.  Pontecorvo did not publish
the report, perhaps because of the times, or because 
Fermi  thought the idea ingenious but not immediately  relevant.

In 1956, using a scintillation counter experiment they had proposed 
three years earlier~\cite{COREA},  Cowan and Reines~\cite{COREB} 
discover electron antineutrinos through the reaction
$\overline \nu_e+p\rightarrow e^+ +n$. Cowan passed away before 1995,
the year Fred Reines was awarded the Nobel Prize for their discovery. 
There emerge two  lessons in neutrino physics: not only is patience
required but also longevity: it took $26$ years from birth to
detection and then another $39$ for the Nobel Committee to recognize 
the achievement! This should encourage   physicists to 
train their children at the earliest age to follow their footsteps at 
the earliest possible age, in order to establish dynasties of neutrino 
physicists.  Perhaps then Nobel prizes will be awarded to scientific families? 

In 1956, it was rumored that  Davis~\cite{DAVISA},  following  Pontecorvo's 
proposal,  had found evidence for neutrinos coming from a pile, and 
Pontecorvo~\cite{PONTB}, influenced by the recent work of   Gell-Mann 
and Pais, theorized that  an antineutrino produced in the Savannah reactor 
could oscillate into a neutrino and be detected. The rumor went
away, but the idea of neutrino oscillations was born; it has remained
with us ever since. 

Neutrinos  give up their secrets very grudgingly: its
helicity was measured in 1958 by M. Goldhaber~\cite{MGOLD},
but it took 40 more years for experimentalists to produce convincing 
evidence  for its mass. The second neutrino, the muon neutrino  is 
detected~\cite{2NEUT} in 1962, (long anticipated by theorists Inou\"e 
and Sakata in 1943~\cite{INSA}). This time things went a bit faster as 
it took only 19 years from theory (1943) to discovery (1962) and 26 years 
to Nobel recognition (1988). 

That same year, Maki, Nakagawa and Sakata~\cite{MANASA} introduce 
two crucial ideas:   neutrino flavors can mix, and their  
 mixing can cause one type of neutrino to oscillate 
into the other (called today flavor oscillation). This is possible 
only if the two neutrino flavors have different masses. 

In 1964, using Bahcall's result~\cite{BAH} of an enhanced capture 
rate of ${^8B}$ neutrinos through an excited state of ${^{37}Ar}$, Davis~\cite{DAVISB}
proposes to search for ${^8B}$ solar neutrinos using a $100,000$ gallon
tank of cleaning fluid deep underground.  Soon after, R. Davis starts 
his epochal experiment at the Homestake mine, marking the
beginning of the solar neutrino watch which continues to this day. In
1968, Davis et al reported~\cite{DAVISC} a deficit in the solar
neutrino flux, 
a result that stands to this day as a truly
remarkable experimental {\it tour de force}. Shortly after, Gribov and
Pontecorvo~\cite{GRIPO} interpreted the deficit as evidence for neutrino oscillations.

In the early 1970's, with the idea of quark-lepton
symmetries~\cite{PASA,GG} suggests that the proton could
be unstable. This brings about the construction of  underground 
 detectors, large enough to monitor many protons, and instrumentalized to 
detect the \v Cerenkov light emitted by its decay products. By the middle 1980's,
several such detectors are in place. They fail to detect proton decay, 
but in a remarkable serendipitous turn of events, 150,000 years earlier, a 
supernova erupted in the large Magellanic Cloud, and in 1987, 
its burst of neutrinos was detected in these detectors! All of 
a sudden, proton decay detectors turn their attention to neutrinos, 
while to this day still waiting for its protons to decay! Today, 
these detectors have shown great success in measuring  
solar and atmospheric neutrinos, culminating in SuperKamiokande's discovery of evidence for neutrino masses.

\section{Standard Model Neutrinos}
The standard model of electro-weak and strong interactions contains three left-handed neutrinos.  The three neutrinos are represented by two-components Weyl spinors, $\nu^{}_{i}$, $i=e,\mu,\tau$, each describing a left-handed fermion (right-handed antifer
mion). As the upper components of weak isodoublets $L^{}_i$, they have $I^{}_{3W}=1/2$, and a unit of the global $i$th lepton number. 

These standard model neutrinos are strictly massless. The only Lorentz scalar made out of these neutrinos is the Majorana mass, of the form
$\nu^{t}_{i}\nu^{}_{j}$; it has the quantum numbers of a weak isotriplet, with third component  $I^{}_{3W}=1$, as well as two units of total lepton number. Thus to generate a Majorana mass term at tree-level, one  needs a Higgs isotriplet with two units 
 of lepton number. Since the standard model Higgs is a 
 weak isodoublet Higgs,  there are no tree-level neutrino masses.

Quantum corrections, on the other hand,  are  not limited to
renormalizable couplings, and it is easy to make a weak isotriplet out
of two isodoublets, yielding the $SU(2)\times U(1)$ invariant
$L^t_i\vec\tau L^{}_j\cdot H^t_{}\vec\tau H$, where $H$ is the Higgs
doublet. As  this term is not invariant under lepton number, it is not be generated in perturbation theory. Thus the
important conclusion: {\it The standard model neutrinos are kept
massless by global chiral lepton number symmetry}. Simply put, {\it neutrino masses offer proof of physics beyond the standard model}.

\section{Neutrino Masses}
Direct experimental limits on neutrino masses are quite impressive, $m_{\nu_e}< 
10~ {\rm eV}$, $m_{\nu_\mu}< 170~ {\rm keV}$, $m_{\nu_\tau}< 18~ {\rm MeV}$~\cite{PDG}, and neutrinos must be extraordinarily light.  Any model that generates neutrino masses must contain a natural mechanism that explains their small value, relative to th
at of their charged counterparts. 

We just just mention one way to generate neutrino masses without new fermions:  add lepton number carrying Higgs fields to the standard model which 
break lepton number explicitly or spontaneously through their interactions.  

Perhaps the simplest way to give neutrinos masses is to introduce for
each one an electroweak singlet Dirac partner, $\overline
N^{}_i$. These appear naturally in the grand unified group
$SO(10)$. Neutrino Dirac masses are generated by the couplings $L^{}_i\overline 
N^{}_j H$ after electroweak breaking. Unfortunately, these Yukawa
couplings yield masses which are like quark and charged lepton masses $m\sim\Delta
I_w=1/2$.  

Based on recent ideas from string theory, it has been
proposed~\cite{HA} that the world of four dimensions is in fact 
a ``brane" immersed in a higher dimensional space. In this view, 
all fields with electroweak quantum numbers live on the brane, 
while standard model singlet fields can live on the ``bulk" as 
well. One such field is the graviton, others could be the right-handed 
neutrinos. Their couplings to the brane are reduced by geometrical 
factors, and the smallness of neutrino masses is due to the 
naturally small coupling between brane and bulk fields.  

In the absence of any credible dynamics for the physics of the bulk, 
and the belief that  {\it ``one neutrino on the brane is worth two 
in the bulk"}, we take  the more conservative approach where  the 
bulk does opens up, but  at much shorter scales. One indication of such 
a scale is that at which the gauge couplings unify, the other is 
given by the value of neutrino masses.
The situation is remedied by introducing
Majorana mass terms $\overline N^{}_i\overline N^{}_j$ for the
right-handed neutrinos. The masses of these new degrees of freedom is
arbitrary, since it has no electroweak quantum numbers, $M\sim\Delta
I_w=0$. If it is much larger than the electroweak scale, the
neutrino masses are suppressed relative to that of their charged
counterparts by the ratio of the electroweak scale to that new scale: 
the mass matrix  (in $3\times 3$ block form) is
\be
\hskip 1in \pmatrix{0& m\cr m&M}\ ,
\ee
leading to one small and one large eigenvalue 
\be
m_\nu~\sim~ m\cdot {m\over M}~\sim~ \left(\Delta I_w={1\over 2}\right)\cdot 
\left({ \Delta I_w={1\over 2}
\over \Delta I_w=0 }\right)\ .\ee 
This seesaw mechanism~\cite{SEESAW} provides a natural
explanation for the smallness of the neutrino masses as long as lepton
number is broken at a large scale $M$. With $M$ around the energy at
which the gauge couplings unify, this yields neutrino masses at or
below the eV region. 

The flavor mixing comes from two different parts, the diagonalization
of the charged lepton Yukawa couplings, and that of the neutrino
masses. From the charged lepton Yukawas, we obtain ${\cal U}_e^{}$, 
the unitary matrix that rotates the
lepton doublets $L^{}_i$. From the neutrino Majorana matrix, we obtain
$\cal U_\nu$, the matrix that diagonalizes the Majorana mass matrix. 
The $6\times 6$ seesaw Majorana matrix can be written in $3\times 3$
block form
\be
{\cal M}={\cal V}_\nu^t ~{\cal D} {\cal V}^{}_\nu\sim\pmatrix {{\cal
U}_\nu&\epsilon {\cal U}^{}_{N\nu}\cr
\epsilon{\cal U}^{t}_{N\nu}&{\cal U}^{}_{N}\cr}\ ,\ee
where $\epsilon$ is the tiny rastio of the electroweak to lepton
number violating scales, and ${\cal D}={\rm diag}(\epsilon^2{\cal D}_\nu, {\cal D}_N)$,
 is a diagonal matrix. ${\cal D}_\nu$ contains the
three neutrino masses, and $\epsilon^2$ is the seesaw suppression. The
weak charged current is then given by
\be
j^+_\mu=e^\dagger_i\sigma_\mu {\cal U}^{ij}_{MNS}\nu_j\ ,\ee
where
\be
{\cal U}^{}_{MNS}={\cal U}^{}_e{\cal U}^\dagger_\nu\ ,\ee
is the matrix first introduced in ref [11], the analog of the CKM
matrix in the quark sector. 

In the seesaw-augmented standard model, this mixing matrix is totally 
arbitrary. It contains, as does the CKM matrix, three rotation angles,
and one CP-violating phase, and also two additional CP-violating phases
which cannot be absorbed in a redefinition of the neutrino
fields, because of their Majorana masses (these extra phases can be measured only in $\Delta {\cal L}=2$ processes). All these additional
parameters await to be  determined by
experiment, although maximal $\nu_\mu-\nu_\tau$ mixing was anticipated long ago~\cite{CASE,HRR} on the basis of grand unified ideas.

\section{Present Experimental Issues}
The best direct limit on the electron neutrino mass come from Tritium
$\beta$ decay, but it does not specify its type, Dirac or Majorana-like. 
An important clue is the absence of neutrinoless double $\beta$ decay, 
which puts a limit on electron lepton number violation. 

Much smaller neutrino masses  can be detected 
through neutrino oscillations. These can be observed using natural 
sources of neutrinos; some are somewhat understood and predictable, such as 
neutrinos produced in cosmic ray secondaries, neutrinos produced 
in the sun; others, such as neutrinos produced in supernovas 
close enough to be detected are much rarer. The second type of 
experiments  monitor neutrinos from reactors, and the 
third type uses accelerator neutrino beams. Below we give 
a brief description of some of these experiments.

\vskip .2cm
\noindent $\bullet$ {\it Atmospheric Neutrinos}

Neutrinos produced in the decay of secondaries from cosmic ray 
collisions with the atmosphere have a definite flavor signature: 
there are twice as many muon like as electron like neutrinos and 
antineutrinos, simply because pions decay all the time into muons. 
It has been known for sometime that this 2:1 ratio differed from 
observation, hinting at a deficit of muon neutrinos. However  
last year  SuperK~\cite{SUPERK} was able to correlate this deficit with the length 
of travel of these neutrinos, and this correlation is the most 
persuasive evidence for muon neutrino oscillations: after birth, 
muon neutrinos do not all make it to the detector as muon neutrinos; 
they oscillate into something else, which in the most conservative 
view, should be either an electron or a tau neutrino. However, a 
nuclear reactor experiment, CHOOZ, rules out the electron neutrino 
as a candidate. Thus there remains two possibilities, the tau neutrino 
or another type of neutrino that does not interact weakly, a 
{\it sterile} neutrino. The latter possibility is being increasingly 
disfavored by a careful analyses of matter effects: it seems 
that muon neutrinos oscillate into tau neutrinos. The oscillation parameters are
\be
(m^2_{\nu_\tau}-m^2_{\nu_\mu})\sim 10^{-3}~{\rm eV}^2\ ;\qquad 
\sin^22\theta_{\nu_\mu-\nu_\tau}\ge .86\ .\ee
Although this epochal result stands on its own, it should be 
confirmed by other experiments. Among these is are experiments 
that  monitor  muon neutrino beams, both at short and long baselines.
\vskip .2cm
\noindent $\bullet$ {\it Solar Neutrinos}

\noindent Starting with the pioneering Homestake experiment, there is clearly 
a deficit in the number of electron neutrinos from the Sun. This has 
now been verified by many experiments, probing different ranges of 
neutrino energies and emission processes. This neutrino deficit 
can be parametrized in three ways

\begin{itemize} 
\item Vacuum oscillations~\cite{GK} of the electron 
neutrino into some other species, sterile or active, can fit the present data, with large mixing angle, $\sin^22\theta_{\nu_
e-\nu_?}\ge .7$, and 
\be
(m^2_{\nu_e}-m^2_{\nu_?})\sim 10^{-10}-10^{-11}~{\rm eV}^2\ .\ee
This possibility implies a seasonal variation of the flux, which the
present data is so far unable to detect.
\item MSW oscillations~\cite{MSW}. In this case, neutrinos produced in the solar
core traverse the sun like a beam with an index of refraction. For a 
large range of parameters, this can result in level crossing region 
inside the sun. There are two distinct cases, according to which the 
level crossing is adiabatic or not. These interpretations yield 
different ranges of fundamental parameters. 

The non-adiabatic layer yields the small angle solution, $\sin^22\theta_{\nu_
e-\nu_?}\ge 2\times 10^{-3}$, and 
\be
(m^2_{\nu_e}-m^2_{\nu_?})\sim 5\times 10^{-6}~{\rm eV}^2 \ .\ee

The adiabatic layer transitions yields the large angle solution, 
$\sin^22\theta_{\nu_
e-\nu_?}\ge 0.65$ with, 
\be
(m^2_{\nu_e}-m^2_{\nu_?})\sim  10^{-4}-10^{-5}~{\rm eV}^2 \ .\ee
This solution implies a detectable day-night asymmetry in the flux.
\end{itemize}

How do we distinguish between these possibilities? Each of these 
implies different distortions of the Boron spectrum from the 
laboratory measurements. In addition, the highest energy solar 
neutrinos may not all come from Boron decay; some are expected 
to be ``hep" neutrinos coming from $p+{^3}He\to {^4}He+e^++\nu_e$.

In their  measurement of the recoil electron spectrum, SuperK 
data show an excess of high end events, which would tend to favor 
vacuum oscillations. They also see a mild day-night asymmetry effect 
which would tend to favor the large angle MSW solution. In short, 
their present data does not allow for any definitive conclusions, 
as it is self-contradictory.

A new solar neutrino detector, the Solar Neutrino Observatory (SNO) 
now coming on-line, should be able to distinguishe between these
scenarios. It contains heavy water, allowing a more 
precise determination of the electron recoil energy, as it involves 
the heavier deuterium. Thus we expect a better resolution of the 
Boron spectrum's distortion. Also, with neutron
detectors in place, SNO will be able to detect all active neutrino species 
through their neutral current interactions. If successfull, this 
will provide a smoking gun test for neutrino oscillations.

 \vskip .2cm
\noindent $\bullet$ {\it Accelerator Oscillations} 

\noindent These have been reported by the LSND 
collaboration~\cite{LSND}, with large angle mixing between 
muon and electron antineutrinos. This result has been partially 
challenged by the KARMEN experiment which sees no such evidence, 
although they cannot rule out the LSND result. This controversy 
will be resolved by an upcoming experiment at FermiLab, called 
MiniBoone. This is a very important issue because, assuming that 
all experiments are correct, the LSND result requires a sterile 
neutrino to explain the other experiments, that is both light 
and mixed with the normal neutrinos. This would require a profound 
rethinking of our ideas about the low energy content of the standard model.


At the end of this Century, there remains several  burning issues in 
neutrino physics that are likely to be soon settled by experiments:

\begin{itemize}
\item Origin of the Solar Neutrino Deficit

This is being addressed by SuperK, in their measurement of the shape 
of the ${^8}B$ spectrum, of day-night asymmetry and of the seasonal 
variation of the neutrino flux. Their reach will soon be improved 
by lowering their threshold energy. 

SNO is  joining the hunt, and is expected to provide a more accurate 
measurement of the Boron flux. Its {\it raison d'\^etre}, however, 
is the ability to measure neutral current interactions. If there 
are no sterile neutrinos, we might have a flavor independent 
measurement of the solar neutrino flux, while measuring at the 
same time the electron neutrino flux!

This experiment will be joined by BOREXINO, designed to measure 
neutrinos from the $^7Be$ capture. These neutrinos are suppressed 
in the small angle MSW solution, which could explain the results 
from the $p-p$ solar neutrino experiments and those that measure the 
Boron neutrinos. 

\item Atmospheric Neutrino Deficit

Here, there are several long baseline experiments to monitor muon 
neutrino beams and corroborate the SuperK results. The first, 
called K2K, already in progress, sends a beam from KEK to SuperK. 
Another, called MINOS, will monitor a FermiLab neutrino beam at 
the Soudan mine, 730 km away. A third experiment under consideration 
would send a CERN beam towards the Gran Sasso laboratory (also about 
730 km away!). Eventually, these experiments hope to detect the 
appearance of a tau neutrino.
\end{itemize}

This brief survey of upcoming experiments in neutrino physics was 
intended to give a flavor of things to come. These measurements will 
not only determine neutrino parameters (masses and mixing angles), 
but will help us answer fundamental questions about the nature of 
neutrinos, especially the possible kinship between leptons and 
quarks. The future of neutrino physics is bright, and with 
much more to come: the  production of intense neutrino beams in 
muon storage rings, and even the detection of the  cosmological 
neutrino background!

\section{Theories}
On the theory side, it must be said that theoretical predictions of lepton hierarchies and
mixings depend very much on hitherto untested theoretical
assumptions. 
In the quark sector, where the bulk of the experimental data resides,
the theoretical origin of quark hierarchies and mixings is a mystery,
although there exits many theories, but none so convincing as to offer a
definitive answer to the community's satisfaction. It is therefore no 
surprise that there are  more theories of lepton
masses and mixings than there are parameters to be measured. Nevertheless, one can formulate the issues in the form of questions:

\begin{itemize}
\item Do the right handed neutrinos have quantum numbers beyond the
standard model?
\item Are quarks and leptons related by grand unified theories?
\item Are quarks and leptons related by anomalies?
\item Are there family symmetries for quarks and leptons?
\end{itemize}

The measured numerical value of
the neutrino mass difference (barring any fortuitous degeneracies), suggests 
through the seesaw mechanism, a mass for the right-handed neutrinos
that is consistent with the scale at which
the gauge couplings unify. Is this just a numerical  coincidence, 
or should we view this as a hint for grand unification?

Grand unified theories, originally proposed as a way to treat
leptons and quarks on the same footing,  imply  symmetries much larger than 
the standard model's. Implementation of these
ideas necessitates a desert and supersymmetry, but also a carefully designed
contingent of Higgs particles to achieve the desired symmetry
breaking. That such models can be built is perhaps more of a testimony
to the cleverness of theorists rather than of Nature's. Indeed with the
advent of string theory, we know that the best features of grand
unified theories can be preserved, as most of the symmetry breaking is
achieved by geometric compactification from higher dimensions~\cite{CANDELAS}.

An alternative point of view is that the vanishing of 
chiral anomalies is necessary for consistent  theories,
and their cancellation is most easily achieved by assembling matter in
representations of anomaly-free groups. Perhaps anomaly cancellation
is more important than group structure.

Below, we present two theoretical frameworks of our work, in which one deduces
the lepton mixing parameters and masses. One is ancient~\cite{HRR},
uses the standard techniques of grand unification, but it had the 
virtue of {\it predicting} the large $\nu_\mu-\nu_\tau$ mixing
observed by SuperKamiokande. The other~\cite{ILR} is more recent, and uses
extra Abelian family symmetries to explain both quark and lepton
hierarchies. It also predicted large $\nu_\mu-\nu_\tau$ mixing, while both
schemes predict small $\nu_e-\nu_\mu$ mixings.

\subsection{A Grand Unified Model}  
The seesaw mechanism was born in the context of the grand unified 
group $SO(10)$, which naturally contains electroweak neutral 
right-handed neutrinos. Each standard model family appears  in 
two irreducible representations of $SU(5)$. However, the predictions of this
theory for Yukawa couplings is not so clear cut, and to reproduce the
known quark and charged lepton hierarchies, a special but simple set of Higgs
particles had to be included. In the simple scheme proposed by Georgi
and Jarlskog~\cite{GJ}, the ratios between the charged leptons and quark masses
is reproduced, albeit not naturally since two Yukawa couplings, not
fixed by group theory, had to be set equal. This motivated us to
generalize~\cite{HRR} their scheme to $SO(10)$, where it is 
(technically) natural, which meant that we had an automatic window
into neutrino masses through the seesaw. The Yukawa couplings were of
the Higgs-heavy,  with ${\bf 126}$ representations, but the attitude at the time 
was ``damn the Higgs torpedoes, and see what happens".  A modern
treatment would include non-renormalizable operators~\cite{BPW}, but
with similar conclusion. The model yielded the mass relations  
\be m_d-m_s=3(m_e-m_\mu)\ ;\qquad m_dm_s=m_em_\mu\ ;\ee
as well as 
\be m_b=m_\tau\ ,\ee
and mixing angles
\be
V_{us}=\tan\theta_c=\sqrt{m_d\over m_s}\ ;\qquad V_{cb}=\sqrt{m_c\over
m_t}\ .\ee
While reproducing the well-known lepton and quark mass hierarchies, it
predicted a long-lived $b$ quark, contrary to the lore of the time.
It also made predictions in the lepton sector, namely
{\bf maximal} $\nu_\tau-\nu_\mu$ mixing, small $\nu_e-\nu_\mu$
mixing of the order of $(m_e/m_\mu)^{1/2}$, and no $\nu_e-\nu_\tau$
mixing. 

The neutral lepton masses came out to be hierarchical, but heavily
dependent on the masses of the right-handed neutrinos. The
 electron neutrino mass came out much lighter
than those of $\nu_\mu$ and $\nu_\tau$. Their numerical values
depended on the top quark mass, which was then supposed to be in the
tens of GeVs!

Given the present knowledge, some of the features are remarkable, such as
the long-lived $b$ quark and the maximal $\nu_\tau-\nu_\mu$
mixing. On the other hand, the actual numerical value of the $b$ lifetime was off
a bit, and the $\nu_e-\nu_\mu$ mixing was too large to reproduce the small angle
MSW solution of the solar neutrino problem. 

The lesson should be that the simplest $SO(10)$ model 
that fits the observed quark and charged lepton hierarchies,
reproduces, at least qualitatively, the maximal mixing found by
SuperK, and predicts small mixing with the electron neutrino~\cite{CASE}

\subsection{A Grand Ununified Model}

There is another way to generate hierarchies, based on adding extra
family symmetries to the standard model, without invoking grand
unification. These types of models address only the Cabibbo
suppression of the Yukawa couplings, and are not as predictive as
specific grand unified models. Still, they predict no Cabibbo
suppression between the muon and tau neutrinos. Below, we present a
pre-SuperK  model~\cite{ILR} with those features. 

The Cabibbo supression is assumed to be an indication of extra
family symmetries in the standard model. The idea is that any standard model-invariant
operator, such as ${\bf Q}_i{\bf \overline d}_jH_d$,  cannot be present
at tree-level if there are additional symmetries under which the
operator is not invariant. Simplest is to assume an Abelian symmetry,
with an electroweak singlet field $\theta$,  as its order parameter.
Then  the interaction
\be
{\bf Q}_i{\bf \overline d}_jH_d\left({\theta\over M}\right)^{n_{ij}}\ee
can appear in the potential as long as the family charges balance under the
new symmetry. As $\theta$ acquires a $vev$, this leads to a
suppression of the Yukawa couplings of the order of $\lambda^{n_{ij}}$
for each matrix element, with 
$\lambda=\theta/M$ identified with  the Cabibbo angle, and
$M$ is the natural cut-off of the effective low energy  theory. 
As a consequence of the charge balance equation
\be X_{if}^{[d]}+n^{}_{ij}X^{}_\theta=0\ ,\ee
the exponents of the suppression are related to the charge of the
standard model-invariant operator~\cite{FN},  the sum of the
charges of the fields that make up the the invariant. 

This simple Ansatz, together with the seesaw mechanism, 
implies that the family structure of the neutrino mass matrix is
determined by the charges of the left-handed lepton doublet fields. 

Each charged lepton Yukawa coupling 
$L_i\overline N_j H_u$, has an extra  charge $X_{L_i}+X_{Nj}+X_{H}$, which
gives the Cabibbo suppression of the $ij$ matrix element. Hence, 
 the orders of magnitude of these couplings can be expressed as 
\be
\pmatrix{\lambda^{l_1}&0&0\cr
0&\lambda^{l_2}&0\cr
0&0&\lambda^{l_3}\cr}{\hat Y}\pmatrix{\lambda^{p_1}&0&0\cr
0&\lambda^{p_2}&0\cr
0&0&\lambda^{p_3}\cr}\ ,\ee
where ${\hat Y}$ is a Yukawa matrix with no Cabibbo
suppressions, $l_i=X_{L_i}/X_\theta$ are the charges of the
left-handed doublets, and 
$p_i=X_{N_i}/X_\theta$, those of the singlets. The first matrix forms half of
the MNS matrix. Similarly, the mass matrix for the right-handed
neutrinos, $\overline N_i\overline N_j$ will be written in the form
\be
\pmatrix{\lambda^{p_1}&0&0\cr
0&\lambda^{p_2}&0\cr
0&0&\lambda^{p_3}\cr}{\cal M}\pmatrix{\lambda^{p_1}&0&0\cr
0&\lambda^{p_2}&0\cr
0&0&\lambda^{p_3}\cr}\ .\ee
The diagonalization of the  seesaw matrix is of the form 
\be 
L_iH_u\overline N_j \left({1\over{{\overline N}~\overline
N}}\right)_{jk}\overline N_kH_uL_l\ ,\ee
from which the Cabibbo suppression matrix from the $\overline N_i$
fields {\it cancels}, leaving us with
\be
 \pmatrix{\lambda^{l_1}&0&0\cr
0&\lambda^{l_2}&0\cr
0&0&\lambda^{l_3}\cr}\hat{\cal M}\pmatrix{\lambda^{l_1}&0&0\cr
0&\lambda^{l_2}&0\cr
0&0&\lambda^{l_3}\cr}\ ,\ee
where $\hat{\cal M}$ is a matrix with no Cabibbo suppressions.  
The Cabibbo structure of the seesaw neutrino matrix is determined
solely by the charges of the lepton doublets! As a result, the Cabibbo
structure of the MNS
mixing matrix is also due entirely to the charges of the three lepton
doublets. This general conclusion depends on the existence of at least
one Abelian family symmetry, which we argue is implied by the observed
structure in the quark sector.

The Wolfenstein parametrization of the CKM matrix~\cite{WOLF}, 
and the Cabibbo structure of the quark mass ratios
\be {m_{u}\over m_t}\sim \lambda ^8\;\;\;{m_c\over m_t}\sim 
\lambda ^4\;\;\; ;
\;\;\; {m_d\over m_b}\sim \lambda ^4\;\;\;{m_s\over m_b}\sim
\lambda^2\ ,\ee
can be reproduced~\cite{ILR,EIR} by a simple {\it family-traceless} charge assignment 
for the three quark families, namely
\be
X_{{\bf Q},{\bf \overline u},{\bf \overline d}} ={\cal B}(2,-1,-1)+
\eta_{{\bf Q},{\bf \overline u},{\bf \overline d}}(1,0,-1)\ ,\ee
where ${\cal B}$ is baryon number, $\eta_{{\bf \overline d}}=0 $, and 
$\eta_{{\bf Q}}=\eta_{{\bf \overline u}}=2$. 
Two striking facts are evident: 
\begin{itemize}
\item the charges of the down quarks, ${\bf \overline d}$, 
associated with the second and third families are the same, 
\item ${\bf Q}$ and ${\bf \overline u}$ have the same value for 
$\eta$.
\end{itemize}
To relate these quark charge assignments to those of the leptons,
we need to inject some more theoretical prejudices. Assume these
 family-traceless charges are gauged, and not anomalous. Then to
cancel anomalies, the leptons must themselves have family charges. 

Anomaly cancellation generically implies group structure. In $SO(10)$, 
baryon number generalizes to ${\cal B}-{\cal L}$, where ${\cal L}$ is total
lepton number, and  in 
$SU(5)$   the fermion assignment is ${\bf\overline 5}={\bf
\overline d}+L$, and ${\bf 10}={\bf Q}+{\bf \overline u}+\overline
e$. Thus anomaly cancellation is easily achieved by
assigning $\eta=0$ to the lepton doublet $L_i$, and $\eta=2$ to the
electron singlet $\overline e_i$, and by generalizing baryon number to
${\cal B}-{\cal L}$, leading to the charges of the three chiral families  
\be
X =({\cal B}-{\cal L})(2,-1,-1)+
\eta_{{\bf Q},{\bf \overline u},{\bf \overline d}}(1,0,-1)\ ,\ee
where now $\eta_{{\bf \overline d}}=\eta_{L}=0 $, and 
$\eta_{{\bf Q}}=\eta_{{\bf \overline u}}=\eta_{\overline e}=2$. 

The charges of the
lepton doublets are simply $X_{L_i}=-(2,-1,-1)$. We have just
argued that these charges determine the Cabibbo structure of the MNS
lepton mixing matrix to be
\be
\hskip .2in
{\cal U}^{}_{MNS}\sim\pmatrix{1&\lambda^3&\lambda^3\cr
\lambda^3&1&1\cr \lambda^3&1&1\cr}\ ,\ee
implying {\it  no Cabibbo suppression in the mixing between
$\nu_\mu$ and $\nu_\tau$}. This is consistent with the SuperK
discovery and  with the small angle MSW~\cite{MSW} solution to the solar
neutrino deficit. One also obtains a much lighter electron neutrino, and
Cabibbo-comparable masses for the muon and tau neutrinos. Notice that
these predictions are  subtly different from those
of grand unification, as they  yield $\nu_e-\nu_\tau$ mixing. 
It also implies a much lighter electron neutrino, and
Cabibbo-comparable masses for the muon and tau neutrinos. 

On the other hand, the scale of the neutrino mass values depend on the family 
trace of the family charge(s). Here we simply quote the results
our model~\cite{ILR}. The masses of the right-handed neutrinos are found to be of the
following orders of magnitude
\be
m_{\overline N_e}\sim M\lambda^{13}\ ;\qquad m_{\overline N_\mu}\sim
m_{\overline N_\tau}\sim M\lambda^7\ ,\ee
where $M$ is the scale of the right-handed neutrino mass terms,
assumed to be the cut-off. The seesaw mass matrix for the three light  neutrinos 
comes out to be 
\be
\hskip .5in
 m^{}_0\pmatrix{a\lambda^6&b\lambda^3&c\lambda^3\cr
b\lambda^3&d&e\cr
c\lambda^3&e&f\cr}\ ,\ee
where we have added for future reference the prefactors $a,b,c,d,e,f$, all of 
order one, and 
\be m_0^{}={v_u^2\over
{M\lambda^3}}\ ,\ee
where $v_u$ is the $vev$ of the Higgs doublet. This matrix has one light eigenvalue
\be
m_{\nu_e}\sim m_0^{}\lambda^6_{}\ .\ee
Without a detailed analysis of the prefactors, the masses of the other 
two neutrinos come out  to be both of 
 order $m_0$. 
The mass difference announced by  superK~\cite{SUPERK}  cannot  be
reproduced without going beyond the model, by taking into account the
prefactors. The two heavier mass
eigenstates and their mixing angle are written in terms of 
\be
x={df-e^2\over (d+f)^2}\ ,\qquad y={d-f\over d+f}\ ,\ee
as
\be {m_{\nu_2}\over m_{\nu_3}}={1-\sqrt{1-4x}\over 1+\sqrt{1-4x}}\
,\qquad \sin^22\theta_{\mu\tau}=1-{y^2\over 1-4x}\ .\ee
If $4x\sim 1$, the two heaviest neutrinos are nearly degenerate. If
$4x\ll 1$, a condition easy to achieve if $d$ and $f$ have the same
sign, we can obtain an adequate split between the two mass
eigenstates. For illustrative purposes, when $0.03<x<0.15$, we find
\be
4.4\times 10^{-6}\le \Delta m^2_{\nu_e-\nu_\mu}\le 10^{-5}~{\rm eV}^2\
  ,\ee
which yields the correct non-adiabatic MSW~\cite{MSW} effect, and
\be
5\times 10^{-4}\le  \Delta m^2_{\nu_\mu-\nu_\tau}\le 5\times
10^{-3}~{\rm eV}^2\ ,\ee
for the atmospheric neutrino effect. These were calculated with a
cut-off, $10^{16}~{\rm GeV}<M<4\times 10^{17}~{\rm GeV}$, and a mixing
angle, $0.9<\sin^22\theta_{\mu-\tau}<1$. This value of the cut-off is 
 compatible not only with the data but also with the gauge
coupling unification scale, a necessary condition for the consistency
of our model, and more generally for  the basic ideas of grand unification.

\section{Outlook}
Presently, neutrino physics is being driven by many
experimental findings that challenge theoretical
expectations. Although all can be explained in terms of neutrino oscillations,
it is unlikely that they are correct in their conclusions: onje must remember that  evidence for neutrino oscillations has often been reported, 
only to either be withdrawn or else contradicted by other experiments. 

The reported anomalies associated with solar neutrinos,
neutrinos produced in cosmic ray cascades~\cite{SUPERK}, and also in low energy
reactions~\cite{LSND},  cannot all be correct without introducing a new
type of neutrino which does not couple to the $Z$ boson, a {\it
sterile}  neutrino~\cite{dienes}.

Small neutrino masses are
naturally generated by the seesaw mechanism, which works because of
the weak interactions of the neutrinos.  A similar mass suppression 
for sterile neutrinos involves new hitherto unknown interactions,
resulting in substantial additions to the standard model, for which
there is no independent evidence. Also, the case for a
heavier cosmological neutrino in helping structure formation may not be as pressing, in
view of the measurements of a small cosmological constant.

Neutrino physics is extremely exciting as it provides  the best opportunities 
for finding and understanding physics beyond the standard model.

\section{Acknowledgments}
I wish to thank Professors Froissart and Vignaud for their kind invitation 
to speak in such a unique setting. This research was supported in part 
by the department of energy under grant DE-FG02-97ER41029.


\begin{thebibliography}{9}

\bibitem{CHAD} J. Chadwick, Verh. d. D. Phys. Ges., {\bf 16}, 383(1914).
\bibitem{ELWO} C. D. Ellis and W. A. Wooster, Proc. Royal Soc. {\bf A117}, 109(1927).
\bibitem{PONTA} B. Pontecorvo, Chalk river Report PD-205, November 1946, unpublished.
\bibitem{COREA} C. L. Cowan and F. Reines, Phys.Rev. {\bf 90}, 492(1953). 
\bibitem{COREB} C.L. Cowan, F. Reines, F.B. Harrison, H.W. Kruse, 
A.D. McGuire, Science {\bf 124}, 103(1956). 
\bibitem{DAVISA} Raymond  Davis Jr., Phys Rev {\bf 97}, 766(1955).
\bibitem{PONTB} B. Pontecorvo, JETP  (USSR) {\bf 34}, 247(1958).
\bibitem{MGOLD} M. Goldhaber, L. Grodzins, A.W. Sunyar, 
Phys.Rev. {\bf 109}, 1015(1958).
\bibitem{2NEUT} G. Danby, J.M. Gaillard, K. Goulianos, L.M. Lederman,
N. Mistry, M. Schwartz, J. Steinberger, Phys.Rev.Lett. {\bf 9}, 36(1962).
\bibitem{INSA} S. Sakata and T. Inou\"e, Prog. Theo. Physics, {\bf 1}, 143(1946).
\bibitem{MANASA} Z. Maki, M. Nakagawa and S. Sakata, Prog. Theo. Physics, {\bf 28}, 247(1962). B. Pontecorvo, Zh. Eksp. Teor. Fiz. {\bf 53}, 1717(1967). 
\bibitem{BAH} J. Bahcall, Phys. Rev. Lett. {\bf 12}, 300(1964).
\bibitem{DAVISB} Raymond  Davis Jr., Phys. Rev. Lett. {\bf 12}, 303(1964).
\bibitem{DAVISC} Raymond  Davis Jr., D. Harmer and K. Hoffman,
Phys. Rev. Lett. {\bf 20}, 1205(1968).
\bibitem{GRIPO} V. Gribov and B. Pontecorvo, Phys. Lett. {\bf B28}, 493(1969).
\bibitem{PASA} J. Pati and A. Salam, Phys. Rev. Lett. {\bf 31}, 661(1973)
\bibitem{GG} H. Georgi and S. L. Glashow, Phys. Rev. Lett. {\bf 32}, 
; H. Fritzsch and P. Minkowski, Annals Phys. {\bf 93}, 193(1975); H. Georgi, in AIP Conference Proceedings no 23, Williamsburg, Va, 1975; F. G\"ursey, P. Ramond and P. Sikivie, Phys. Lett. {\bf 60B}, 177(1976)
\bibitem{PDG} Particle Data Group, R. M. Barnett {\it et al.}, Phys
Rev {\bf D54}, 1(1996).
\bibitem{HA}
N. Arkani-Hamed, S. Dimopoulos, Gia Dvali {\it
Phys.Lett.} {\bf B429}, 263(1998); E. Dudas, K. Dienes, and
T. Gherghetta, {\it Phys.Lett.} {\bf B436}, 55(1998)
\bibitem{SEESAW} M. Gell-Mann, P. Ramond, and R. Slansky in Sanibel
Talk, CALT-68-709, Feb 1979 (unpublished), and in {\it Supergravity} (North Holland,
Amsterdam 1979). T. Yanagida, in {\it Proceedings of the Workshop on
Unified Theory and Baryon Number of the Universe}, KEK, Japan, 1979.
\bibitem{CASE} The Case for Neutrino Oscillations, P. Ramond, 
Proceedings of the Los Alamos Neutrino Workshop, LA-9358-C, June 1981.
\bibitem{HRR}  J. A. Harvey, P. Ramond and 
D. B. Reiss, Nucl. Phys. {\bf B199}, 223(1982)

\bibitem{SUPERK} Super-Kamiokande Collaboration, Phys. Rev. Lett. 
{\bf 81}, 1562(1998)

\bibitem{GK} S. L. Glashow and L. M. Krauss, Phys.Lett. {\bf 190B}, 199(1987)
\bibitem{MSW} L. Wolfenstein, Phys. Rev. D17, 2369 (1978); S. Mikheyev and A. Yu Smirnov, Nuovo Cim. {\bf 9C}, 17 (1986).
\bibitem{LSND} C. Athanassopoulos {\it et al.}, Phys. Rev. Lett. {\bf
75}, 2560(1995); {\bf 77}, 3082(1996); nucl-ex/9706006.

\bibitem{CANDELAS} P. Candelas, G. Horowitz, A. Strominger and
E. Witten, Nucl. Phys. {\bf B258}, 46(1985) 

\bibitem{ILR} N. Irges, S. Lavignac and P. Ramond, Phys. Rev. {\bf
D58}, 035003(1998).
\bibitem{GJ} H. Georgi and C. Jarlskog, Phys. Lett. {\bf B86}, 297(1979)

\bibitem{BPW} K. S. Babu, J. Pati and F. Wilczek, hep-ph/9812538.

\bibitem{FN} C.~Froggatt and H.~B.~Nielsen Nucl. Phys. B147 (1979) 277; P. Ramond, R.G. Roberts and G.G. Ross, Nucl. Phys. B406 (1993)
\bibitem{WOLF} L. Wolfenstein, Phys. Rev. Lett. {\bf 51}, 1945(1983).
\bibitem{EIR} J. Elwood, N., Irges, and P. Ramond, Phys. Rev. Lett. {\bf 81},  5064(1998)
 
\bibitem{dienes} For an intriguing possibility see,  E. Dudas, K. Dienes, and
T. Gherghetta, hep-ph/9811428.

\end{thebibliography}
\end{document}